\documentclass{webofc}
\usepackage[varg]{txfonts}   
\usepackage[right]{lineno}
\usepackage[export]{adjustbox}
\usepackage{orcidlink}

\begin{document}

\title{Data Movement Model for the Vera C. Rubin Observatory}


\author{    \firstname{Fabio} \lastname{Hernandez} \inst{1} \orcidlink{0000-0001-7203-2552} \and
    \firstname{Mark~G.} \lastname{Beckett} \inst{2} \orcidlink{0000-0003-3623-9753} \and
    \firstname{Andrew} \lastname{Hanushevsky} \inst{3} \and
    \firstname{Tim} \lastname{Jenness} \inst{4} \orcidlink{0000-0001-5982-167X} \and
    \firstname{Kian-Tat} \lastname{Lim} \inst{3} \orcidlink{0000-0002-6338-6516} \and
    \firstname{Peter} \lastname{Love} \inst{5} \and
    \firstname{Timothy} \lastname{Noble} \inst{6} \and
    \firstname{Stephen} \lastname{Pietrowicz} \inst{7} \orcidlink{0000-0002-2158-6480} \and
    \firstname{Wei} \lastname{Yang} \inst{3}}
\institute{
CNRS, CC-IN2P3, 21 avenue Pierre de Coubertin, CS70202, F-69627 Villeurbanne cedex, France \and
Institute for Astronomy, University of Edinburgh,  Royal Observatory, Blackford Hill, Edinburgh EH9 3HJ, UK \and
SLAC National Accelerator Laboratory, 2575 Sand Hill Rd., Menlo Park, CA 94025, USA \and
Vera C.\ Rubin Observatory Project Office, 950 N.\ Cherry Ave., Tucson, AZ  85719, USA \and
Lancaster University, Lancaster, UK \and
Science and Technology Facilities Council, Rutherford Appleton Laboratory, Harwell, UK \and
NCSA, University of Illinois at Urbana-Champaign, 1205 W.\ Clark St., Urbana, IL 61801, USA
}

\abstract{
The sky images captured nightly by the camera on the Vera C. Rubin Observatory’s telescope will be processed across facilities on three continents. Data acquisition will occur at the observatory's location on Cerro Pachón in the Andes mountains of Chile. A first copy of the raw image data set is stored at the summit and immediately transmitted via dedicated network links to the archive center within the US Data Facility at SLAC National Accelerator Laboratory in California, USA and from there to two European facilities for processing and archiving purposes. Data products resulting from periodic processing campaigns of the entire set of images collected since the beginning of the survey are made available to the scientific community in the form of data releases.

In this paper we present an overall view of how we leverage the tools selected for managing the movement of data among the Rubin processing and serving facilities, including Rucio and FTS. We also present the tools we developed to integrate Rucio's data model and Rubin's Data Butler, the software abstraction layer that mediates all access to storage by pipeline tasks that implement science algorithms.
}

\maketitle

\section{Introduction}
\label{introduction}
The Vera C. Rubin Observatory's mission is to explore the universe by conducting the \textit{Legacy Survey of Space and Time} (LSST), the largest-ever sky survey with an unprecedented wide-field imaging system. The observatory aims to capture deep, high-resolution images of the night sky, mapping the cosmos to investigate fundamental questions in astrophysics \cite{Ivezic:2019}.

\begin{figure}[h]
\includegraphics[width=\textwidth]{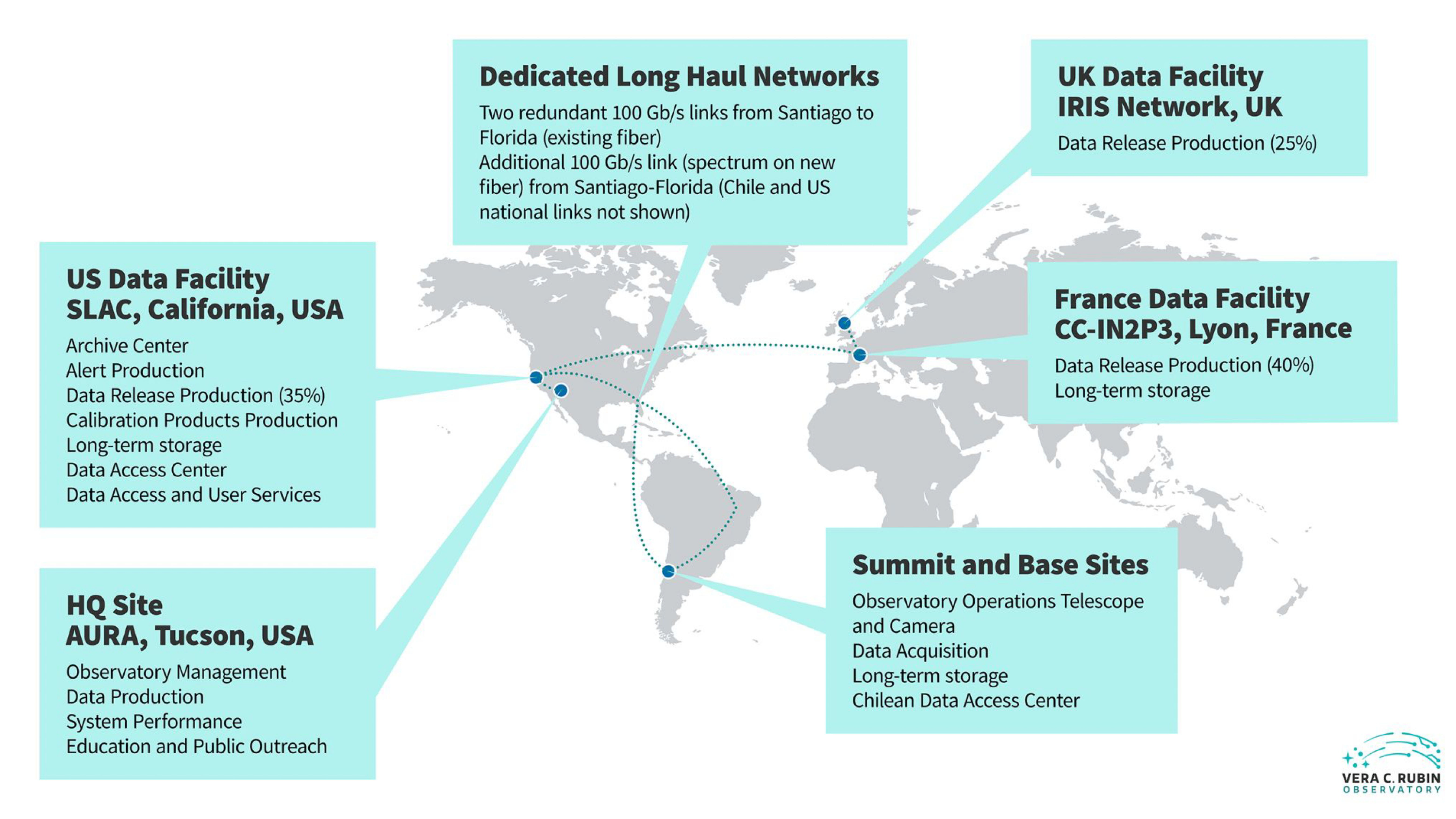}
\caption{Raw images flow from the Summit Site, where the telescope is located in Chile, to the Base Site and then to the Archive Center within the US Data Facility through long haul network links specifically deployed for the needs of the Observatory. Data is transferred from the Archive Center to the European Data Facilities for processing and archiving. The US, UK and France Data Facilities collectively provide the computational capacity for processing the images taken by the Observatory for the duration of the survey. The Observatory headquarters are located in Tucson, USA.}
\label{fig:data-facilities}
\end{figure}

The sky images captured nightly by the observatory's 3.2-gigapixel camera covering the wavelength range 320--1050 nm will be processed across facilities on three continents. Data acquisition will occur at the observatory's location on Cerro Pachón in the Andes mountains of Chile. A first copy of the raw image data is stored at the summit and immediately transmitted via dedicated network links to the Archive Center within the US Data Facility at SLAC National Accelerator Laboratory in California, USA (see Fig.\ \ref{fig:data-facilities}). There, every observing night, an image processing campaign uses the Prompt Processing pipelines to generate data products about transients, variables and moving objects, which results in distribution of alerts within 60 seconds of image readout \cite{LSE-163,melissa_graham_2022_7011229}. After an embargo period of 80 hours, the full raw image dataset is transferred to the France Data Facility, where a third copy is maintained, and a partial dataset is transferred to the UK Data Facility.

Over its 10-year operational period, starting late 2025, annual processing campaigns will be jointly conducted by the France, UK and US Data Facilities on all images collected since the beginning of the survey. Sophisticated algorithms will extract measurements of celestial objects from these images, producing science-ready images and catalogs. Data products resulting from these processing campaigns will be sent to the Archive Center for integration into a consistent Data Release, which will be made available to the scientific community through Data Access Centers in the US and Chile, as well as Independent Data Access Centers (IDACs) elsewhere.

The remainder of this paper is structured as follows. We present in section \ref{section-data-movement-use-cases} the main data movement use cases we need to satisfy and in section \ref{section-data-movement-tools} the tools that have been selected or developed and how they are composed to implement solutions to those use cases.

\section{Data movement use cases}
\label{section-data-movement-use-cases}

A dataset of about 5 PB of new image data will be recorded by the instrument every year, for a total of 50 PB of raw data accumulated over the duration of the survey. Processing the input dataset for the purpose of producing a data release generates approximately ten times the size of the input dataset, including intermediate datasets not part of the the published release.

This section presents three distinct use cases for moving data among the data facilities used by the Rubin Observatory.

\subsection{From summit to archive}
\label{summit-to-archive}

The data acquisition system stores each exposure as a set of approximately 200 files, one per sensor on the camera focal plane. Once an exposure is recorded at the summit site, its constituent files are transferred in parallel to an object store at the archive center via the S3 protocol \cite{s3}. To optimize these transfers over the international network linking the summit to the SLAC archive site, we employ specialized network connection pooling, keep-alive mechanisms, and TCP tuning.

Given that raw images undergo prompt processing for transient object detection and alert generation, the target end-to-end latency for transferring a single exposure—including data compression and other overheads—is set to seven seconds for four gigabytes of compressed data.

Ancillary data (e.g., telemetry, specialized databases) are replicated to the archive center using native protocols to avoid translation steps that can add latency and complexity. Additionally, a small number of certified calibration files are transferred infrequently from the archive to the summit and other locations.

\subsection{From archive to processing facilities and back}
\label{summit-to-data-facilities}

Annual processing of the entire image dataset recorded since the beginning of the survey is carried out across three facilities: the US Data Facility, hosted at SLAC National Accelerator Laboratory in California, USA\footnote{\url{https://www.slac.stanford.edu}}, the France Data Facility, hosted by the IN2P3 computing center (CC-IN2P3) in Lyon, France\footnote{\url{https://cc.in2p3.fr}}, and the UK Data Facility, operated by the LSST:UK consortium\footnote{\url{https://www.lsst.ac.uk}}.

Raw image data is replicated from the US to the European facilities. Both the US and France data facilities store a complete copy of the raw image dataset. The UK facility receives the raw images corresponding to the spatial region assigned to it for processing. Data movement between these sites is facilitated by ESnet\footnote{\url{https://es.net}}, which handles transatlantic data transport; GEANT\footnote{\url{https://geant.org}}, which connects European sites; and the national research and education networks, JANET\footnote{\url{https://www.jisc.ac.uk/janet}} (UK) and RENATER\footnote{\url{https://renater.fr}} (France).

The entire set of final data products, along with selected intermediate products from each campaign, is replicated from the facility where they are generated to the archive center. There, they are consolidated and incorporated into a new data release, which is delivered annually to the science community for analysis \cite{10.1051_epjconf_202429501042}.

The LSST Science Pipelines are the software tools developed by Rubin Observatory to process the survey data \cite{PSTN-019}. They include advanced image processing algorithms and supporting middleware. A central component of this middleware is the Rubin Data Butler, an abstraction layer that mediates access to the data required by, or generated through, the pipelines \cite{2022SPIE12189E..11J}. The Data Butler retrieves data from persistent storage (using appropriate protocols and data formats) based on queries specified by scientifically relevant identifiers (rather than file paths), and delivers the data as in-memory Python objects to the pipelines. It also persists the in-memory objects generated by the science algorithms. Crucially, the Butler manages the location of all files within the data store, recording their locations and relationships in a relational database. Together, the file registry and the storage system where files are located constitute a \emph{repository}.

Since a given Butler repository is aware only of the files present at a single facility, files replicated between facilities need to be placed in the repository's data store at the location expected by the Butler. Upon reception, replicated files are ingested into the receiving facility's local Butler repository, making them available for the processing pipelines.

\subsection{From archive to data access centers}
\label{summit-to-data-access-centers}

Annually released data products must be distributed to approximately 15 to 20 IDACs across the Americas, Europe, and Asia-Pacific regions, where scientific analysis will be conducted. These distribution campaigns will be centrally coordinated by Rubin to ensure timely delivery of data releases to all analysis centers. The goal is to distribute the multi-petabyte datasets to the IDACs in a tiered manner, with some centers receiving data directly from the Archive Center and then sending the data on to other IDACs, thereby reducing the load on the Archive Center \cite{RTN-086}.

\section{Data movement tools}
\label{section-data-movement-tools}

Several software tools are employed to implement the use cases outlined in the previous section. CERN's Rucio \cite{rucio2019} and its companion FTS \cite{FTS} manage the movement of files between the archive site and data facilities, as well as from the archive to the data access centers. In addition, Rubin-specific tools have been developed to register files and automate actions when replicated files arrive at their destination. These tools and their usage are described in the following subsections.

Rubin Observatory operates a dedicated instance of Rucio, configured to transfer files between the Rucio Storage Elements (RSEs) at each facility. These storage endpoints support a data movement protocol that Rucio utilizes to transport data across them. The US and UK data facilities use XrootD \cite{xrootd}, while the France data facility uses dCache \cite{dCache}. All of these systems expose the webDAV protocol \cite{webdav}, an extension of HTTP \cite{http1.1}. Data is transferred securely across sites using confidential channels built on top of secure HTTP.

Each processing facility exposes at least two storage elements to Rubin's Rucio, both typically served by the same storage system. One RSE is used for storing input data required for processing (e.g., raw images, calibration data, reference catalogs) and the other endpoint for storing the products generated by the image processing pipelines \cite{DMTN-213}. Data stored by the input data RSE is protected against modification and removal and a full copy is archived to tape at both the US and France data facilities. Data products stored in the products RSE are less sensitive as they can be regenerated and even some of them may be deleted after a processing campaign is complete. Logically separating the storage endpoints for inputs and data products provides the ability to prioritize transfers for raw data over data products, in addition to providing an extra layer of protection for the critical raw data.

All RSEs are configured to use the identity logical-to-physical filename mapping. This configuration ensures that the file pathnames are preserved relative to the Butler repository's datastore location, which is critical for proper file replication to the destination where the Butler expects to find them. 

\subsection{Registration of files to replicate}
\label{registration}

To perform replication, we create Rucio Datasets, each composed of a set of files that are already in their appropriate locations at the source RSE. Upon registration, preconfigured Rucio subscriptions trigger the actual file movement to the destination facility, in accordance with the defined replication rules. Rucio delegates the execution of file transfers to FTS, which then instructs the storage endpoints at the facilities to move the data, typically by requesting the destination facility to pull the data from the source facility. The use of Datasets allows grouping of related files, use of subscription patterns applied to spatially defined Dataset names to associate spatial regions with RSEs, and a clear way to know when all related files have been generated (via Dataset closure) and replicated.

Rubin has developed the tool \texttt{rucio\_register}\footnote{\url{https://github.com/lsst/rucio_register}}, which allows for the selection of existing files from a Butler repository based on specified criteria. The tool attaches Rubin-specific metadata to these files and registers them into one or more Rucio Datasets. The metadata, encoded as a JSON record, contains a minimal set of information extracted from the source Butler repository. This ensures that replicated files can be ingested properly into the local Butler repository at the destination facility.

We use \texttt{rucio\_register} to make known to Rucio only the files that need replication between any two of the Rubin data facilities. As a result, Rubin's instance of Rucio is aware only of files replicated across processing facilities. Files that are local to each facility and not subject to replication remain known to that facility's Butler registry database and datastore but are not registered in Rucio. Since the US Data Facility gets a complete copy of all final data products, files that are not replicated are, by definition, intermediates in the calculations that are not required to be persisted once the final data products are generated and can then be deleted. Considering that the number of intermediate files may be ten times greater than the number of final data products, storing information about intermediates in Rucio is not cost-effective.

\begin{figure}[h]
\includegraphics[width=0.7\textwidth, center]{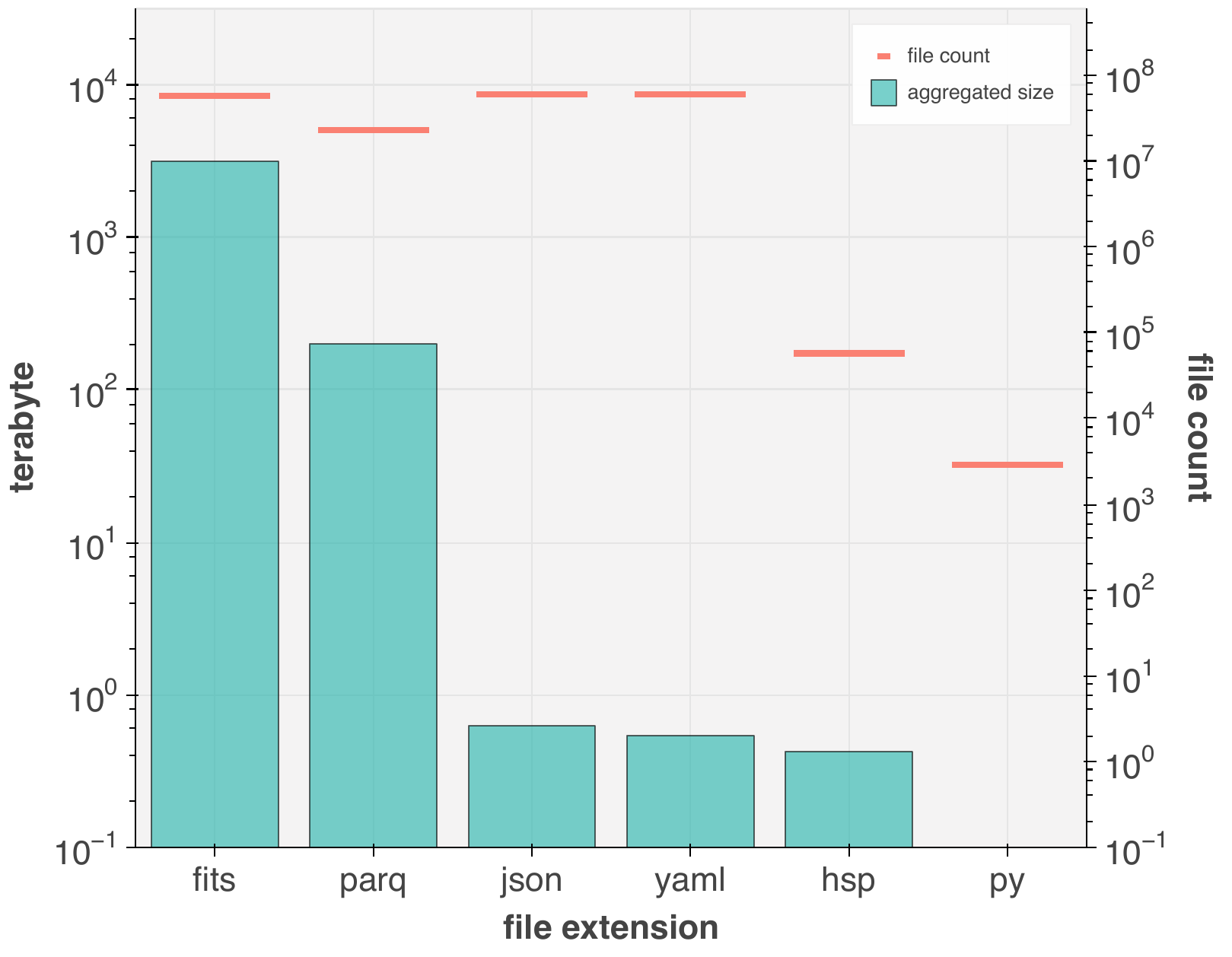}
\caption{Number of files and total file sizes from a data preview processing run.}
\label{fig:filecount}
\end{figure}

The pipeline processing generates many ancillary files in addition to pixel data.
A data preview processing run \cite{10.1051/epjconf/20242950404} demonstrated that the number of JSON and YAML files is approximately of the same scale as the number of FITS and Parquet data files (see Fig.\ \ref{fig:filecount}).
Given that the ancillary files are significantly smaller (sometimes a few kB per file) this can lead to very large file transfer overheads.
To mitigate this problem we have modified the Butler infrastructure to allow the small files from a single processing run to be combined into one or more Zip files.
These Zip files contain the Butler metadata necessary to allow the Butler to retrieve individual files whilst making a single file available to Rucio.

\subsection{Ingestion at reception}
\label{ingestion}

FTS notifies Rucio about the completion of individual file transfers. Rubin's \texttt{HermesK}\footnote{\url{https://github.com/lsst-dm/ctrl_rucio_ingest}}, which is a modification of Rucio's Hermes daemon, filters messages and uses Kafka as a mechanism to signal to the destination Rubin facility that a new file was replicated and to take appropriate actions. Kafka was selected as a reliable message bus used for other purposes within the Rubin project (see e.g., \cite{2024SPIE13101E..1MF,2024SPIE13101E..18R}). Its ordering guarantees are not strictly necessary in this application, but the ability to scale to multiple consumers may be needed as the number of files increases.

Messages distributed through Rubin's Kafka control-plane include Rubin-specific metadata. Those messages are received by Rubin's \texttt{ingestd}\footnote{\url{https://github.com/lsst-dm/ctrl_ingestd}}, a daemon running at each destination facility responsible for ingesting newly replicated files into the local Butler repository.

\begin{figure}[h]
\includegraphics[width=0.9\textwidth, center]{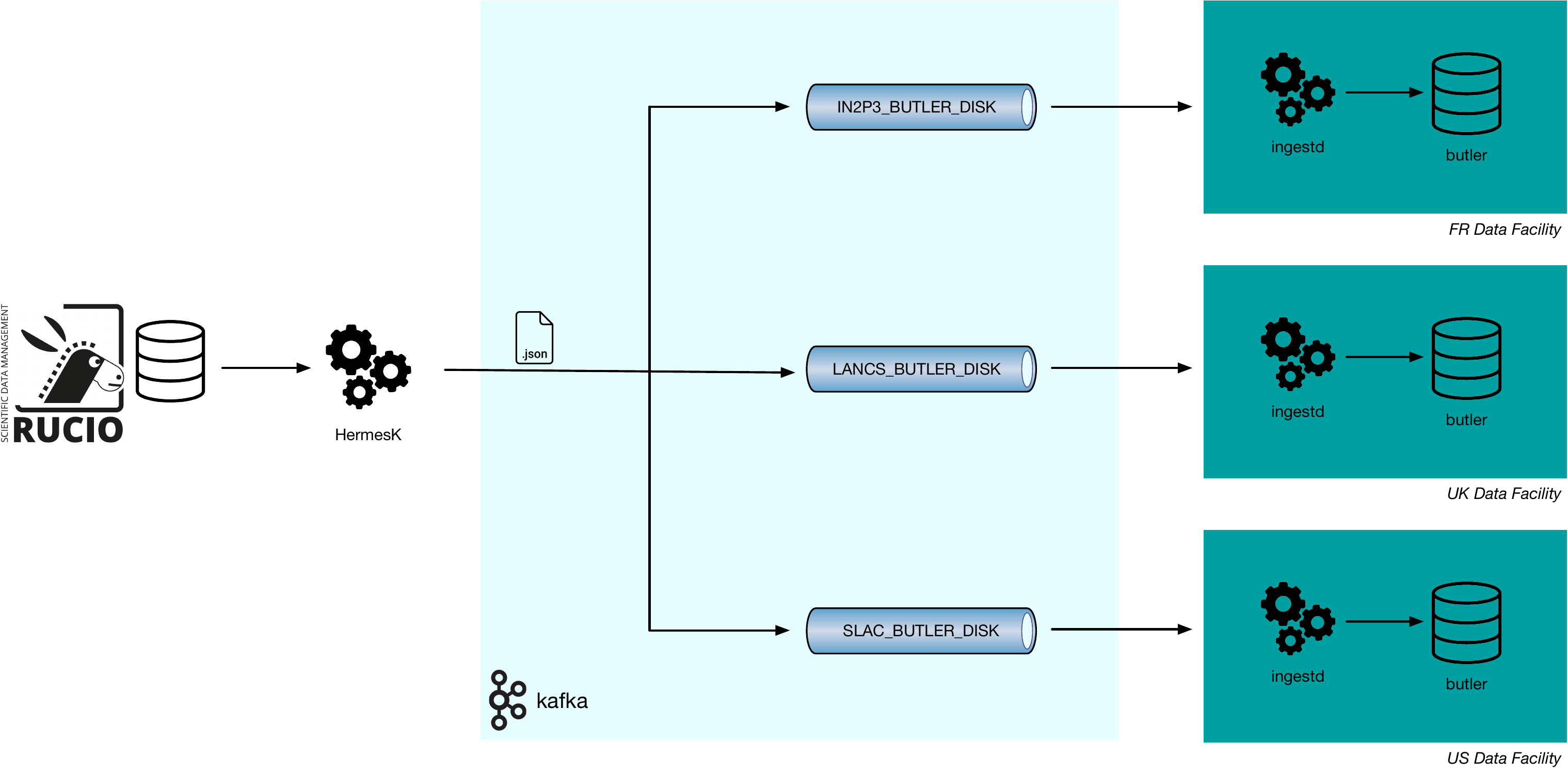}
\caption{\texttt{HermesK} emits notifications about successful file transfers via Kafka topics named after the destination RSE. At the receiving facility \texttt{ingestd} monitors those notifications and ingests the newly received file into the local Butler repository. The JSON-encoded, Rubin-specific metadata associated to the file when it was first registered into Rucio contains the details needed for ingestion.}
\label{fig:kafka-control-plane}
\end{figure}

Each facility only receives notifications about files successfully replicated to the storage endpoints it operates. This is achieved by following a simple convention: the name of Kafka topic the notification is sent to is identical to the name of the Rucio storage element. Each facility's \texttt{ingestd} is configured to only monitor Kafka messages specifically targeted to the facility's RSEs (see Fig.\ \ref{fig:kafka-control-plane}).

In a complex distributed system such as this, having stateless daemons such as \texttt{ingestd}, idempotent transactions such as ingestion of file batches, and triggering off known synchronization points such as replication acknowledgement helps ensure a consistent, if conservative, view of the available data across multiple sites.

\section{Summary}
\label{summary}
We presented several use cases for the movement of data among the facilities participating in the processing of Rubin Observatory data, the tools used to implement solutions to satisfy those uses cases as well as the tools Rubin has developed for integrating Rubin-specific software components to more generic software systems for large-scale inter-site data transfer.

\section{Acknowledgments}

This material is based upon work supported in part by the National Science Foundation through Cooperative Agreement AST-1258333 and Cooperative Support Agreement AST-1202910 managed by the Association of Universities for Research in Astronomy (AURA), and the Department of Energy under Contract No. DE-AC02-76SF00515 with the SLAC National Accelerator Laboratory managed by Stanford University. Additional Rubin Observatory funding comes from private donations, grants to universities, and in-kind support from LSSTC Institutional Members.

This work has been supported by the UK Science and Technology Facilities Council (STFC) funding for UK participation in LSST, through grants ST/X001334/1 and ST/Y003004/1.

This work has been supported by the French National Institute of Nuclear and Particle Physics (IN2P3) through dedicated funding provided by the National Center for Scientific Research (CNRS).

\bibliography{references}

\end{document}